\documentclass[a4paper]{jpconf}
\usepackage{graphicx}
\usepackage{amsmath,amssymb}
\usepackage{epsfig}
\usepackage{dcolumn}
\usepackage{bm}
\usepackage{theorem}
\usepackage{amsmath,amssymb}

\newcommand{\qed}{\nobreak \ifvmode \relax \else
\ifdim\lastskip<1.5em \hskip-\lastskip \hskip1.5em plus0em
minus0.5em \fi \nobreak \vrule height0.75em width0.5em
depth0.25em\fi}

\input{epsf}

\newcommand{\bmp}{\begin{minipage}}
\newcommand{\emp}{\end{minipage}}
\newcommand{\bc}{\begin{center}}
\newcommand{\ec}{\end{center}}
\newcommand{\beq}{\begin{equation}}
\newcommand{\eneq}{\end{equation}}

\begin{document}
\title{Mini-bandstructure tailoring in $\pi$-conjugated periodic block copolymers employing the envelope crystalline-orbital method}
\author{C. A. Mujica-Mart\'inez and J. C. Arce}
\address{Departamento de Qu\'imica, Universidad del Valle, A. A. 25360, Cali, Colombia}
\ead{cesarmujica@gmail.com, jularce@univalle.edu.co
}
\begin{abstract}
A strategy for the systematic design of polymeric superlattices with tailor-made mini-bandgaps and carrier mini-effective masses is described and computationally implemented by means of an envelope crystalline-orbital method, which is a straightforward adaptation for molecules of the envelope-function approximation widely used in solid-state physics. Such strategy relies on the construction of $\pi$-conjugated periodic block copolymers from well-characterized parent polymers, in such a way that the above-mentioned electronic parameters can be predicted from the lengths of the blocks. Illustrative calculations for prototypical (PPP$_{x}$-PDA$_{y}$)$_{n}$ superlattices demonstrate the plausibility of the strategy and the advantages of the computational implementation employed.
\end{abstract}
\noindent
To appear  in {International Journal of Quantum Chemistry}

\section{Introduction}
\label{sec:introduction}
A solid-state superlattice is a periodic arrangement of many nanometric layers of two, or more, different semiconductor materials \cite{inorganicSL,organicSL}. The ionization energies and electron affinities of these materials are such that the effective potential-energy profiles for electrons and holes along the direction of growth take the forms of periodic arrays of quantum wells. The barriers between the wells are thin enough so that their otherwise isolated single-particle levels become resonantly coupled by quantum tunneling. Such coupling mixes these levels, giving rise to so-called energy mini-bands, in an analogous fashion to the appearance of energy bands in a crystal. In this sense, a superlattice behaves as an artificial crystal with a much larger period (of the order of $10$ nm) than any real crystal, and, consequently, with a much smaller (mini-)Brillouin zone in reciprocal space. The mini-bandwidths and, consequently, the carrier mini-effective masses are determined by the thickness of the barriers, whereas the mini-bandgaps are determined by the depths and widths of the wells. Therefore, the mini-band structure of a superlattice can be engineered by controlling these parameters \cite{yu-cardona}.\\

\noindent Crystalline heterostructures such as this can be manufactured from inorganic \cite{inorganicSL} and organic \cite{organicSL} materials, employing sophisticated, and costly, epitaxial crystal-growth methods, like molecular-beam epitaxy and chemical or physical vapor deposition techniques \cite{yu-cardona}. These methods afford so good a control over the fabrication that the miniband structures can be tailored for specific electronic and optoelectronic applications \cite{SLapplications}.\\

\noindent This report deals with a different kind of superlattice, which is constituted by a quasi-linear $\pi$-conjugated 
alternating block copolymer \cite{previousworks}. Now, the effective potential-energy profiles for carriers assume the forms 
of periodic arrays of tunnel-coupled quantum dots. (To conform to the current usage in the semiconductor literature, it is more appropriate to employ the term `quantum dot' than `quantum well' in this case, since in the former the carriers are confined in the three directions, whereas in the latter they are confined in one direction only.) Hence, this organic heterostructure can be considered as a molecular onedimensional analog of the aforementioned solid-state layered superlattice.\\

\noindent Molecular organic heterostructures such as this present the following advantages over their solid-state counterparts: First, in principle, it is possible to construct a virtually unlimited variety of them, due, in turn, to the virtually unlimited number of monomers available, whereas the number of parent crystalline materials available is limited \cite{yu-cardona}. Second, the ``interface'' between two blocks is a chemical carbon-carbon bond, in contrast to the interface between two crystalline layers where it is required that the materials exhibit similar lattice constants \cite{yu-cardona}, which notably limits the variety of heterostructures that can be obtained. Third, organic copolymers \cite{organiccopolymers} and heterostructures \cite{exptalorgstruc} can be obtained through chemical-synthesis methodologies, whose implementation is easier and cheaper than the combination of epitaxial crystal-growth and lithographic methods commonly used for the manufacture of crystalline heterostructures.\\

\noindent In the light of these observations, the general objective of this contribution is to point out that molecular 
quantum-dot superlattices, constituted by quasi-linear $\pi$-conjugated alternating block copolymers, afford an attractive 
alternative to solid-state quantum-well superlattices for the design of materials with engineered bandgaps and conductivities. Specifically, it shall be shown that, by appropriately selecting the parent polymers and defining the block lengths, the mini-bandgaps and carrier mini-effective masses of these molecular organic heterostructures can be tailored at will. Hence, this strategy is complementary to the one presented in the previous contribution to this issue \cite{complementarypaper}, where the blocks are so short that the copolymer does not behave like a superlattice but instead like a new polymer, whose bandgap and effective masses can be controlled by the mole fractions of the parent monomers. (It should be clarified that these regular copolymers have also been called `superlattices' by other authors \cite{regularcopolymers}; however, this is not recommendable since these systems do not exhibit mini-bands).\\

\noindent The implementation of this design strategy will be purely computational. In particular, firstly the parent polymers will be characterized by means of quantum-chemical electronic structure and bandstructure methods. Secondly, once the block copolymers are designed and assembled, their mini-bandstructures will be determined employing an envelope crystalline-orbital approximation \cite{perdomo,mujicathesis,inpreparation}, which is conceptually more attractive and computationally more economical for this task than any of the conventional atomistic quantum-chemical methods employed thus far for copolymers \cite{previousworks,regularcopolymers}. This method is a straightforward adaptation for molecules of the envelope-function, or effective-mass, approximation widely employed in solid-state physics \cite{yu-cardona,davies}.

\section{Design Strategy}
\label{sec:design}

The design strategy consists of the following steps: (1) Choice of a pair $A$-$B$ of parent polymers with an appropriate alignment of their valence-band tops, or highest unoccupied crystal orbitals (HOCO's), and conduction-band bottoms, or lowest unoccupied crystal orbitals (LUCO's). (2) Design of the repeat unit (quantum dot) of the superlattice, to guarantee that it possess the desired number of bound levels and range of separations between them. (3) Assembly of the superlattice, or alternating block copolymer, from the repeat unit of the previous step. (4) Determination of the mini-bandgaps and electron and hole mini-effective masses of the superlattice, as functions of its structural parameters. (5) Construction of calibration curves of such quantities as functions of these parameters, from which the same quantities for an arbitrary superlattice can be extracted. The meanings of these steps are now explained.\\

\begin{figure}[h]
\begin{center}
\includegraphics[width=0.7\textwidth]{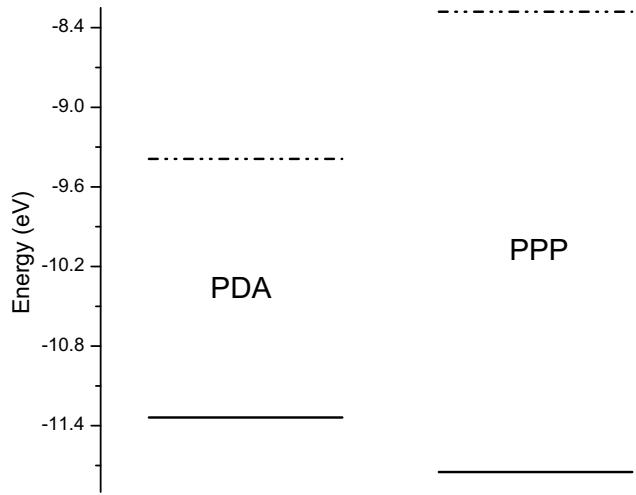}
\end{center}
\caption{\label{MOpolymersaligment} \footnotesize{Alignment of the PDA and PPP frontier CO's (band edges). Solid lines: 
HOCO's (valence-band tops). Dashed lines: LUCO's (conduction-band bottoms).}} 
\end{figure}

\noindent To illustrate the strategy, for step (1) poly-(\textit{p}-phenylene) (PPP) and poly-diacetylene (PDA) were selected as parent polymers for the construction of quasi-linear block copolymers. Fig. \ref{MOpolymersaligment} displays the HOCO and LUCO energies of these polymers, calculated using the methodology indicated in Section \ref{sec:methodology}. This alignment of the frontier CO's indicates that `type I' heterostructures \cite{yu-cardona,davies} can be designed from these parent polymers (see below).

\vspace{0.3cm}

\begin{figure}[h]
\begin{center}
\includegraphics[width=0.7\textwidth]{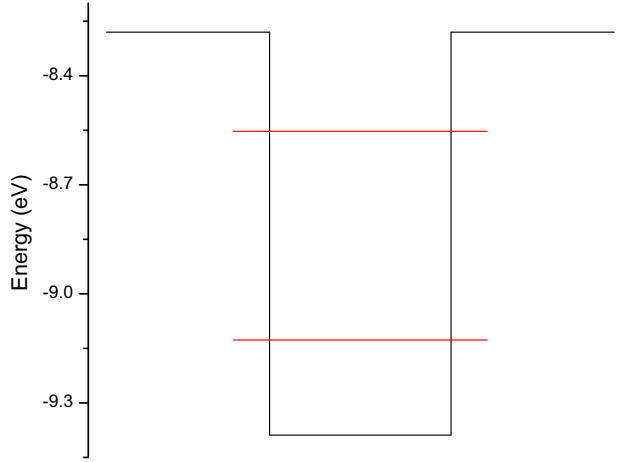}
\end{center}
\caption{\label{electronicstruture-8-7-8} \footnotesize{Top: molecular structure of the quantum dot PPP$_8$-PDA$_7$-PPP$_8$. Bottom: Electron potential profile and energy levels calculated by means of the Extended H\"uckel method. The barriers associated with the ``heterojunctions'' between the molecular frame and the left and right vacua are not shown.}} 
\end{figure}

\noindent For step (2), the three-block molecular heterostructure PPP/PDA/PPP is chosen as repeat unit for the superlattice. According to the envelope molecular-orbital (EMO) theory \cite{yu-cardona,perdomo,mujicathesis,inpreparation,davies}, the heterojunctions between blocks can be considered as abrupt, so that the effective potential-energy profiles along the longitudinal direction for electrons and holes within the molecular frame take the forms of finite square wells, as illustrated in Fig. \ref{electronicstruture-8-7-8}. On the other hand, the ``heterojunctions'' between the molecular frame and the left and right vacua can be represented by infinitely high abrupt potential barriers with good approximation \cite{perdomo,mujicathesis,inpreparation} (not shown in Fig. \ref{electronicstruture-8-7-8}). Since the carriers are confined in the three directions, this heterostructure is zero-dimensional and should be called `quantum dot' instead of `quantum well'.

\vspace{0.3cm}

\noindent Now, in each region the electron and hole quantum states are governed simply by corresponding one-dimensional effective-mass Schr\"odinger equations \cite{yu-cardona,inpreparation,davies,harrison}
\begin{equation}\label{Schrodinger-eq}
  \left[ - \frac{\hslash^2}{2 m^*_i(x)} \frac{d^2}{dx^2} + V_i(x) \right] \psi_i(x) = \varepsilon_i \psi_i(x)
\end{equation}

\noindent where $x$ is the coordinate along the longitudinal direction of the molecule, and $V_i(x)$, $m^*_i(x)$, $\varepsilon_i$ and $\psi_i(x)$ are the potential energy, (position-dependent) parabolic effective mass, eigenenergy and EMO of the carrier ($i =$ electron or hole), respectively. For electrons (holes), the height of the barriers is given by the difference, or `offset', between the energies of the LUCO's (HOCO's) of the parent polymers. The EMO approximation comes in handy because the bound electron and hole states can be estimated by the familiar particle-in-a-box expression \cite{davies,harrison}
\begin{equation}\label{particle-in-a-box}
 \varepsilon_i \approx E_i + \frac{\pi^2 \hslash^2 n_i^2}{2 m_i^* l_w^2},
\end{equation}

\noindent where $\varepsilon_i$ is measured from the edge of its respective band, $E_i$, $m_i^*$ is the carrier parabolic effective mass of PDA and $l_w$ is the width of the potential well.\\

\noindent That this heterostructure is of type I means that the potential wells for electrons and holes are aligned (`straddling'). This implies that the superlattice can exhibit mini-bands both for electrons and holes. This report considers electron mini-bands only, the hole mini-bands being analogous. Moreover, attention will be focused on the mini-bands arising from two parent electron levels bound to the intramolecular well. With the help of Eq. (\ref{particle-in-a-box}), it was predicted that the quantum dot PPP$_8$-PDA$_7$-PPP$_8$ contains two electron bound states, which were subsequently calculated more accurately employing the quantum-chemical methodology indicated in Section \ref{sec:methodology}. The chemical structure, electron potential profile and energy levels of this quantum dot are shown in Fig. \ref{electronicstruture-8-7-8}. The length of the molecule is $10.2$ nm.\\

\begin{figure}[h]
\begin{center}
\includegraphics[width=1.0\textwidth]{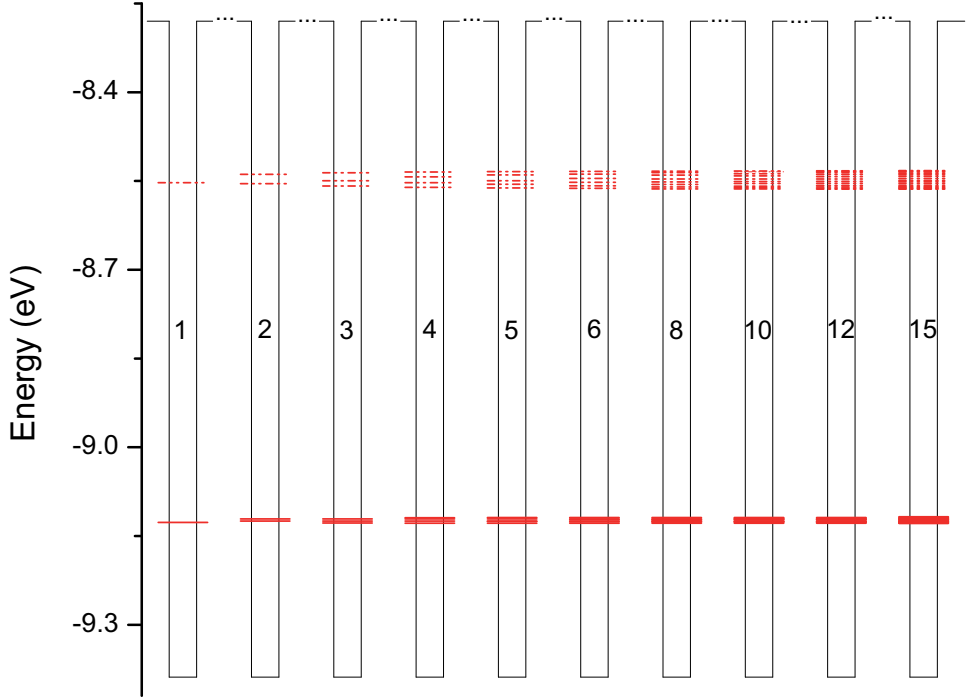}
\end{center}
\caption{\label{evolutionofenergylevels} \footnotesize{Evolution of the energy levels in the periodic arrays of increasing lengths PPP$_8$(-PDA$_7$-PPP$_8$)$_{n}$ calculated by means of the Extended H\"uckel method. For reference, the levels of each array are drawn inside the potential profile of the parent quantum dot.}} 
\end{figure}

\noindent Before proceeding to the superlattice, it is instructive to examine the evolution of the energy levels as the number of coupled quantum dots increases. Fig. \ref{evolutionofenergylevels} illustrates such an evolution for periodic arrays of quantum dots with molecular structures PPP$_8$(-PDA$_7$-PPP$_8$)$_{n}$, determined employing the methodology indicated in Section \ref{sec:methodology}. As expected, it is observed that a manifold of $n$ levels emerges from each parent level. In addition it is seen that for $n=15$ the two manifolds already appear as quasi-continuous mini-bands. Naturally, in the hypothetical limit $n \rightarrow \infty$ the manifolds will become continuous 
mini-bands \cite{yu-cardona,davies,harrison}.

\vspace{0.3cm}

\begin{figure}[h]
\begin{center}
\includegraphics[width=0.8\textwidth]{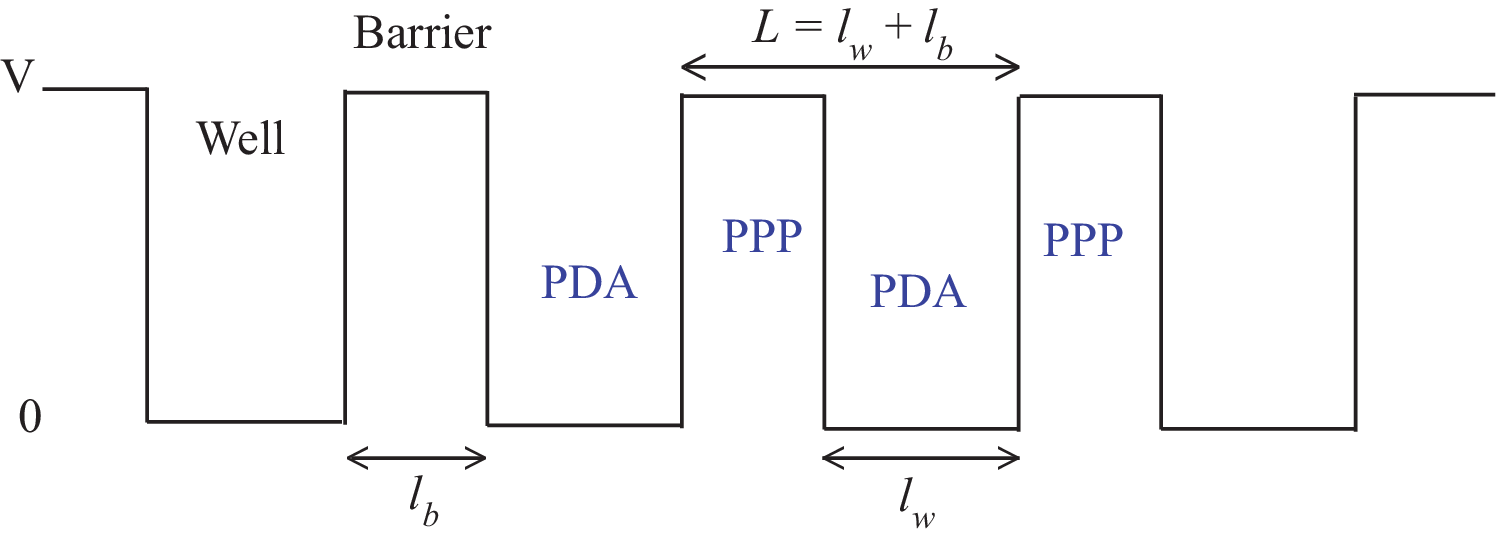}
\end{center}
\caption{\label{KPpotential} \footnotesize{Potential profile for electrons in the superlattice (PDA$_7$-PPP$_8$)$_{n}$. The energy is
measured from the LUCO of PDA.}} 
\end{figure}

\noindent Within the framework of the EMO theory, step (3), the assembly of the superlattice from the quantum dot just designed, is straightforward. The effective potential profile for electrons in the superlattice takes the form of a Kronig-Penney model \cite{harrison}, i.e. a periodic array of square wells, as displayed in Fig. \ref{KPpotential}. It is evident that two parameters, the length of the well ($l_w$) and the length of the barrier ($l_b$), are available for mini-bandstructure tailoring purposes. $l_w$ largely determines the energies of the parent discrete levels, allowing to perform `level engineering'. Consequently, this parameter will also control the locations of the mini-bands and partly the (mini-)gap between them, permitting to carry out `bandgap engineering'. On the other hand, $l_b$ determines the size of the overlap between adjacent parent EMO's. Hence, it will control the mini-bandwidths, and, consequently, the mini-effective masses, allowing to perform `effective-mass engineering'.

\vspace{0.5cm}
\vspace{0.3cm}

\noindent The Kronig-Penney model already helps to explain qualitatively the behavior of the energy levels displayed in Fig. \ref{evolutionofenergylevels}. Specifically, the first manifold is narrower (denser) than the second because the overlap between adjacent EMO's of the ground levels of the individual quantum dots is smaller than for the excited ones. This is so because the height of the barrier is effectively smaller for the excited levels, making the tunneling probability between them larger.\\

\noindent The methodology for the determination of the mini-bandgaps and electron and hole mini-effective masses as functions of $l_w$ and $l_b$, entailed in steps (4) and (5), will be explained in the next section.

\section{Computational Methodology}
\label{sec:methodology}

The bandstructures of the parent polymers PPP and PDA were calculated employing the semiempirical methodology of the previous contribution to this issue \cite{complementarypaper}, which entails a combination of the Austin Model 1 \cite{AM1}, for geometry optimizations, with the Extended H\"uckel method \cite{EHM}, for single-point electronic-structure calculations. Such methodology yields accurate HOCO and LUCO energies, but does not guarantee accurate bandwidths or, consequently, effective masses. Therefore, the PPP and PDA electron and hole parabolic effective masses were determined by means of an accurate real-space technique, based on interpolations of HOMO and LUMO data for oligomers of increasing sizes, which is explained elsewhere \cite{perdomo,mujicathesis,inpreparation}. The electronic structures of the PPP/PDA/PPP quantum dot and the finite superlattice PPP$_8$(-PDA$_7$-PPP$_8$)$_{n}$ were calculated using a methodology analogous to the one employed for copolymers in ref. \cite{complementarypaper}, which yields somewhat less accurate HOMO and LUMO values than the HOCO and LUCO ones produced by the above-mentioned method employed for the parent polymers.\\

\noindent The mini-bandstructure of the superlattice (PDA$_x$-PPP$_y$)$_{n}$ can be determined employing periodic boundary conditions and any quantum-chemical electronic structure method (see the previous paper of this issue \cite{complementarypaper} and references therein). However, for construction of calibration curves as functions of $x$ and $y$, this would be very tedious and time-consuming. Furthermore, these methods produce a lot of atomistic information that is not required for the present purposes. On the other hand, the EMO method, employing the Kronig-Penney model, permits the calculation of the mini-bandstructure with reasonable accuracy, as long as the blocks are not too short \cite{inpreparation}, and very little computational effort \cite{perdomo,mujicathesis,inpreparation}, as is now explained.\\

\noindent In a finite superlattice, the carriers tend to be more localized towards the center of the structure, although states localized at the chain ends (Tamm states) may appear \cite{harrison}. On the other hand, in an infinite superlattice the carriers are always equally likely to be found in any of the wells. Consequently, the envelope orbitals are periodic and satisfy Bloch's theorem \cite{yu-cardona,davies}
\begin{equation}\label{Bloch}
 \psi(x,k) = u(x,k) \exp(ikx).
\end{equation}

\noindent where $u(x,k)$ is a background function that possesses the same periodicity as the potential and $k$ is the wavenumber along the longitudinal direction $x$. The first mini-Brillouin zone corresponds to $-\pi/L \leq k \leq \pi/L$, where $L = l_w + l_b$ is the period of the potential. With the exception of the possible appearance of Tamm states, an infinite superlattice is an adequate model for a long, finite superlattice. Thus, for simplicity, that model will be considered from this point on, and its envelope functions will be called ``envelope crystalline orbitals'' (ECO's) to differentiate them from the localized EMO's.\\

\noindent Within the well and barrier regions the ECO's can be written as \cite{harrison}
\begin{eqnarray}
\label{envelopewell}
 \psi_w & = & A \exp(i k_w x) + B \exp(-i k_w x)\\
\label{envelopebarrier}
 \psi_b & = & C \exp(i k_b x) + D \exp(-i k_b x)
\end{eqnarray}

\noindent respectively, where $k_w = \sqrt{2 m_w^* \varepsilon / \hslash^2}$ and $k_b = \sqrt{2 m_b^* (\varepsilon-V) / \hslash^2}$, 
with $m^*_r$ ($r = w,b$) the effective mass of the charge carrier inside the region $r$.\\

\noindent To guarantee good behavior of the ECO's, these expressions must be matched across the two PPP-PDA heterojunctions present in the unit cell, enforcing the BenDaniel-Duke boundary conditions \cite{yu-cardona,davies,harrison,bendaniel-duke} that both $\psi(x)$ and $\frac{1}{m^*(x)} \frac{d\psi_r}{dx}$ be continuous at these points. Moreover, the periodicity condition \cite{harrison}
\begin{equation}\label{periodicity}
 \psi(x + L) = \psi(x) \exp(ikL).
\end{equation}

\noindent implied by Bloch's theorem (Eq. (\ref{Bloch})) must also be taken into account. (Notice that the Bloch wave number $k$ gives the phase change of the ECO's from one unit cell to the next, whereas the wave number $k_r$ ($r = w,b$) governs the variations of these functions within each region). Employing these conditions, four coupled homogeneous equations are obtained, which, after some algebra, yield the transcendental equation \cite{harrison}
\begin{equation}\label{dispersionrelation}
 \cos(k_w l_w) \cosh(\kappa l_b) - \sin(k_w l_w) \sinh(\kappa l_b) \left( \frac{m_b^2 k_w^2 - m_w^2 \kappa^2}{2 m_w m_b k_w \kappa} \right) = \cos(kL)
\end{equation}

\noindent where $\kappa = \sqrt{2 m_b^* (V-\varepsilon) / \hslash^2}$. The values of the effective masses that appear in this equation are taken from ref. \cite{perdomo} and are displayed in Table \ref{table}. The superlattice dispersion curves $\varepsilon (k)$ were obtained from graphical solutions of Eq. (\ref{dispersionrelation}).

\section{Results and Discussion}
\label{sec:results}

\begin{figure}[h]
\begin{center}
\includegraphics[width=0.75\textwidth]{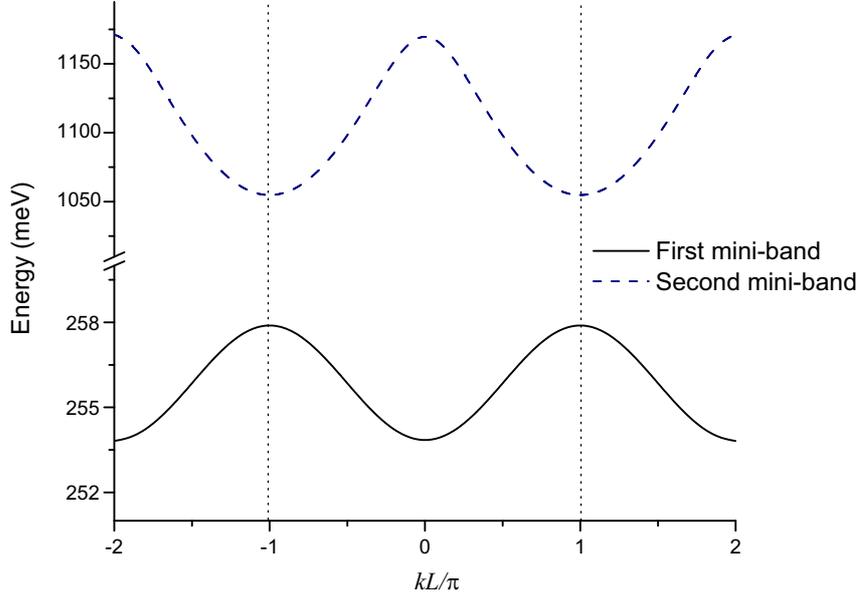}
\end{center}
\caption{\label{minibands} \footnotesize{Mini-bandstructure of the infinite superlattice (PPP$_8$-PDA$_7$)$_{n \rightarrow \infty}$, calculated employing the ECO approximation and the Kronig-Penney model. The energy is measured from the LUCO of PDA.}} 
\end{figure}

Fig. \ref{minibands} displays the two lowest-energy solutions of Eq. (\ref{dispersionrelation}), in the extended mini-zone scheme, for the prototypical superlattice (PPP$_8$-PDA$_7$)$_{n \rightarrow \infty}$. It is worth noticing at the outset that the top and bottom of the first and second mini-bands, respectively, occur at the edge of the mini-zone. This implies that this kind of structure can be used for the design of inter-mini-band lasers, which are high-power sources of mid-infrared radiation \cite{minibandlaser}.

\vspace{0.3cm}

\noindent Analogously to bulk crystals, a gap between two mini-bands is defined as the difference between the top of the lower-lying one and the bottom of the higher-lying one. By the same token, a mini-bandwidth is defined as $\Delta \equiv |\varepsilon(k=0)-\varepsilon(k=\pi/L)|$. The calculated mini-bandgap and the bandwidths of the first and second mini-bands of this superlattice are presented in Table \ref{table}. The corresponding results of the electronic-structure calculation for the small superlattice PPP$_8$(-PDA$_7$-PPP$_8$)$_{15}$ shown in Fig. \ref{evolutionofenergylevels} are also presented in Table \ref{table}. These two sets of results should not be compared directly, since, as mentioned in Section \ref{sec:methodology}, the semiempirical electronic-structure methodology employed for this heterostructure produces somewhat inaccurate HOMO-LUMO gaps and unreliable bandwidths (and, consequently, effective masses). That is probably why the mini-bandgap of PPP$_8$(-PDA$_7$-PPP$_8$)$_{15}$ turns out to be smaller than the one of PPP$_8$(-PDA$_7$-PPP$_8$)$_{n \rightarrow \infty }$ by about 30$\%$, and the first and second mini-bandwidths of the former are off by factors of three and four, respectively, with respect to the ones of the latter. However, the semi-quantitative agreement between these results is indicative of the plausibility of the ECO method. For comparison, the bandgaps and conduction-bandwidths of the parent polymers are also reported in Table \ref{table}. It is observed that the superlattice mini-bandgaps fall in the mid-IR, whereas the parent polymer bandgaps fall in the UV-Vis.\\

\noindent Within the parabolic approximation, the effective mass of a particle with a dispersion relation $\varepsilon (k)$ can be obtained from $\frac{1}{m^*} = \frac{d^2\varepsilon (k)}{dk^2}$ \cite{yu-cardona,davies,harrison}. The mini-effective masses of the electron around $k=0$, relevant for direct transitions, for the first and second mini-bands, extracted by means of this equation from the dispersions of Fig. \ref{minibands}, are reported in Table \ref{table}. It is interesting to compare these values with the electron effective masses of PDA and PPP (see Table \ref{table}) \cite{perdomo}. It is seen that the effective mass of the first mini-band is much larger than the ``bulk'' values, whereas the absolute value of the effective mass of the second mini-band is less than the bulk values. Since a carrier's mobility is inversely proportional to its effective mass \cite{yu-cardona}, this implies that electrons in the second mini-band will have a greater response, and in the opposite sense, to an electric field than in the bulk, whereas the contrary occurs for the first mini-band. From now on, for the effective mass of the second mini-band only the absolute value will be reported.\\

\begin{table}
\caption{\label{table} Bandgaps ($E_g$), conduction-band widths ($\Delta_2$) and electron effective masses ($m_2^*$) of the parent polymers; and mini-bandgaps ($E_g$), mini-bandwidths ($\Delta_1$, $\Delta_2$) and mini-effective masses ($m_1^*$,$m_2^*$) of the prototypical finite and infinite superlattices. $m_0$ is the electron mass.}
\begin{center}
\begin{tabular}{cccccc}
\hline
\hline
                  System                    &    $E_g$/meV    &  $\Delta_1$/meV  &  $\Delta_2$/meV  &  $m^*_1$/$m_0$  &   $m^*_2$/$m_0$  \\
\hline
                    PPP                     &   1.95x10$^3$   &                  &   6.43x10$^3$    &                 &       0.034      \\
                    PDA                     &   3.47x10$^3$   &                  &   7.05x10$^3$    &                 &       0.102      \\
      PPP$_8$(-PDA$_7$-PPP$_8$)$_{15}$      &       554       &       11.2       &       31.2       &                 &                  \\
 (PPP$_8$-PDA$_7$)$_{n \rightarrow \infty}$ &       796       &       4.11       &        117       &      0.818      &      -0.015      \\
\hline
\hline
\end{tabular}
\end{center}
\end{table}
\vspace{0.3cm}

\begin{figure}[h]
\begin{center}
\includegraphics[width=0.75\textwidth]{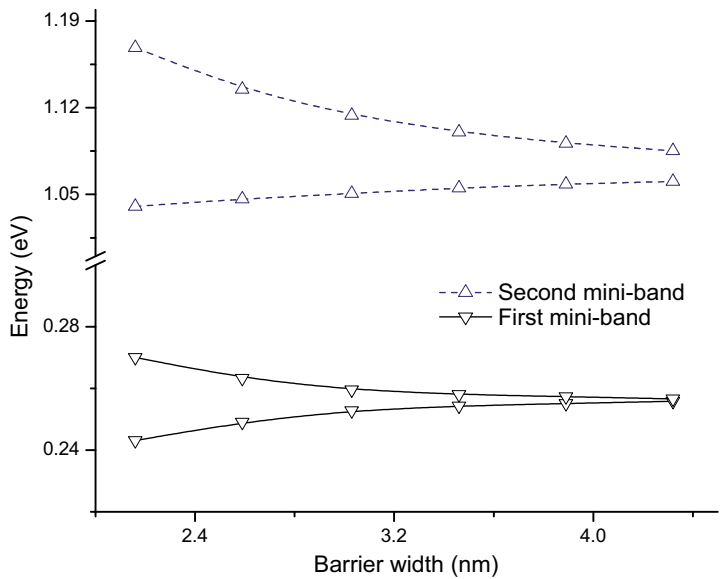}
\end{center}
\caption{\label{minibandsedges} \footnotesize{Allowed energies (region between the same kind of line) of the infinite superlattices 
(PPP$_x$-PDA$_7$)$_{n \rightarrow \infty}$, where $x$ was varied from $5$ to $10$. The energy is measured from the LUCO of PDA. The oligomer PPP$_8$ corresponds to a barrier width of $\sim 3.5$ nm.}} 
\end{figure}

\noindent Let us now examine the effect of increasing the number of monomers in the barrier (PPP) regions over the mini-bandstructure for the 
superlattices (PPP$_x$-PDA$_7$)$_{n \rightarrow \infty}$, with $x = 5,..,10$, according to the ECO method employing the Kronig-Penney model. 
Fig. \ref{minibandsedges} shows the behavior of the extrema of each mini-band, where the lines joining the data, as in all the forthcoming figures, are obtained by interpolation. 

\vspace{0.3cm}

\begin{figure}[h]
\begin{center}
\includegraphics[width=0.49\textwidth]{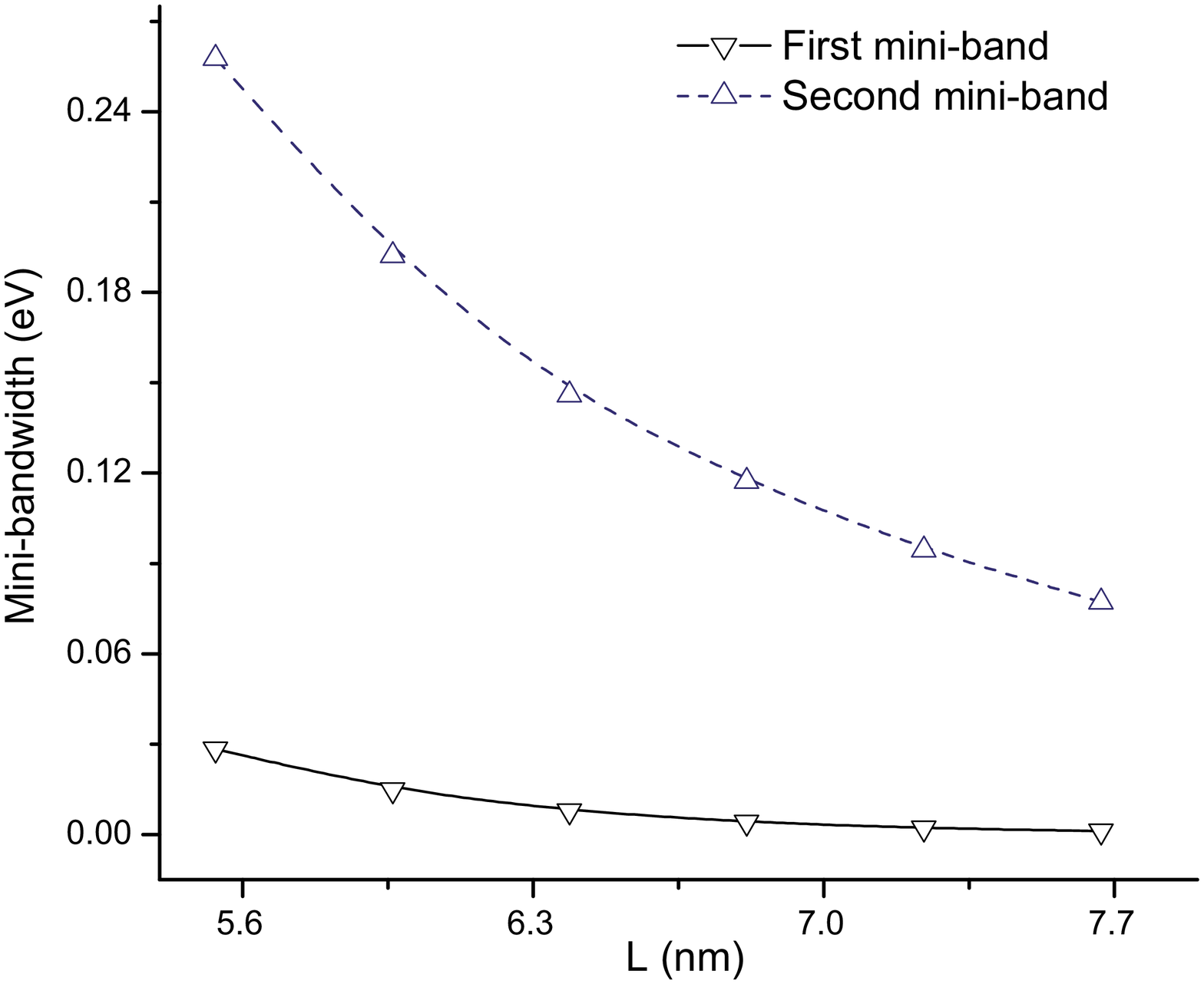} \,
\includegraphics[width=0.49\textwidth]{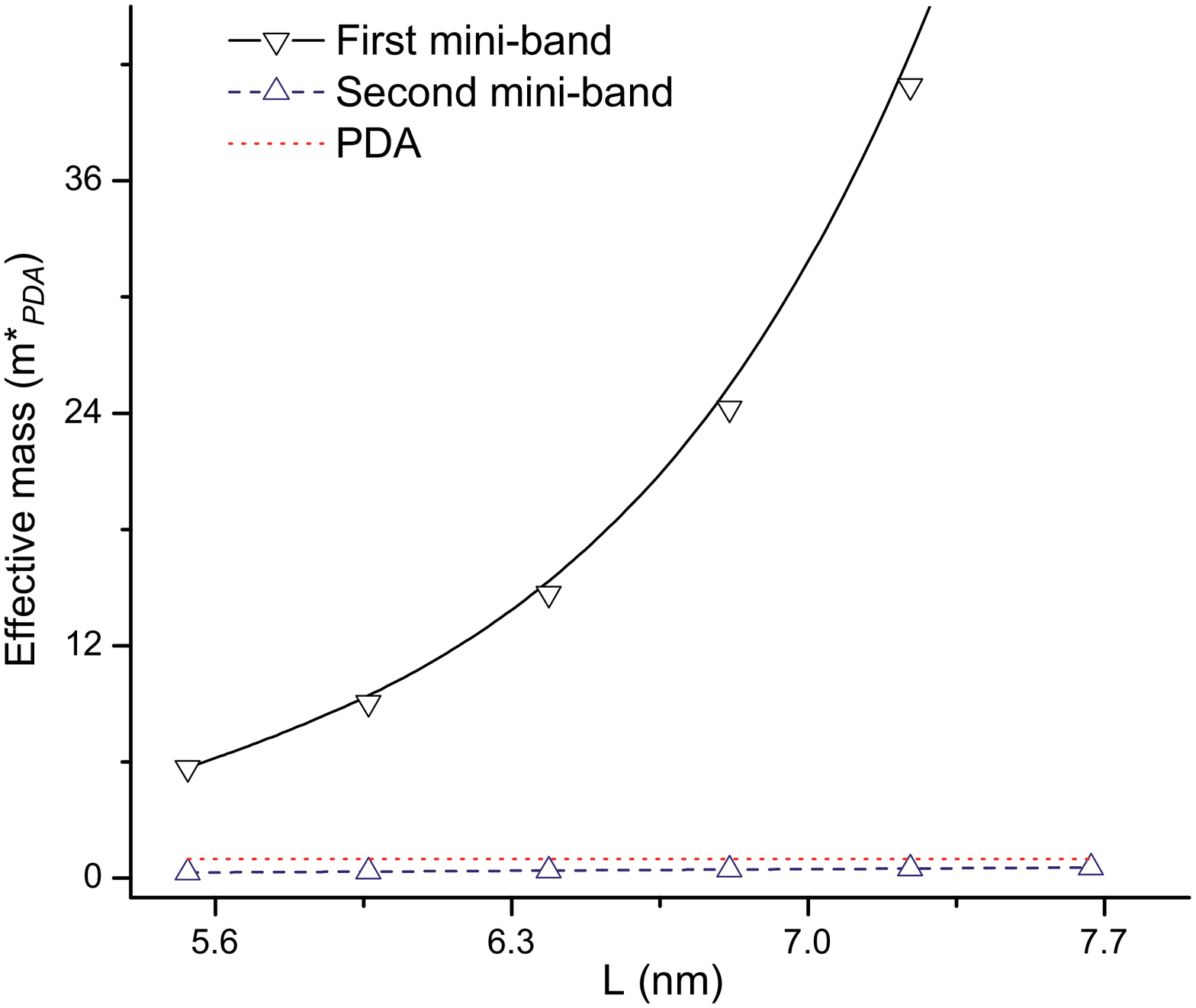}
\end{center}
\caption{\label{complete-x-7-x} \footnotesize{Infinite superlattices (PPP$_x$-PDA$_7$)$_{n \rightarrow \infty}$. Left: mini-bandwidths. Right: mini-effective masses in units of the PDA electron effective mass (dotted line).}} 
\end{figure}

\noindent As expected, it is seen that these get closer together as the barrier thickness increases. For a very thick barrier all the states within the mini-band will finally merge towards the corresponding level of the single well. This is a crossover from a superlattice to an array of non-interacting identical quantum dots. The left part of Fig. \ref{complete-x-7-x} displays the behavior of the mini-bandwiths, where it is observed that the convergence toward zero is monotonic.

\vspace{0.5cm}
\vspace{0.5cm}

\noindent The right part of Fig. \ref{complete-x-7-x} shows the behavior of the mini-effective masses, where it can be appreciated that, as compared to the one of the first mini-band, the effective mass of the second mini-band is largely insensitive to the barrier thickness, remaining below the bulk PDA value. This behavior is due to the proximity of the second mini-band to the edge of the barrier, which endows these states with a much stronger PPP character than the states of the first mini-band, making them much less sensitive to variations in the thickness of the barrier.\\

\begin{figure}[h]
\begin{center}
\includegraphics[width=0.49\textwidth]{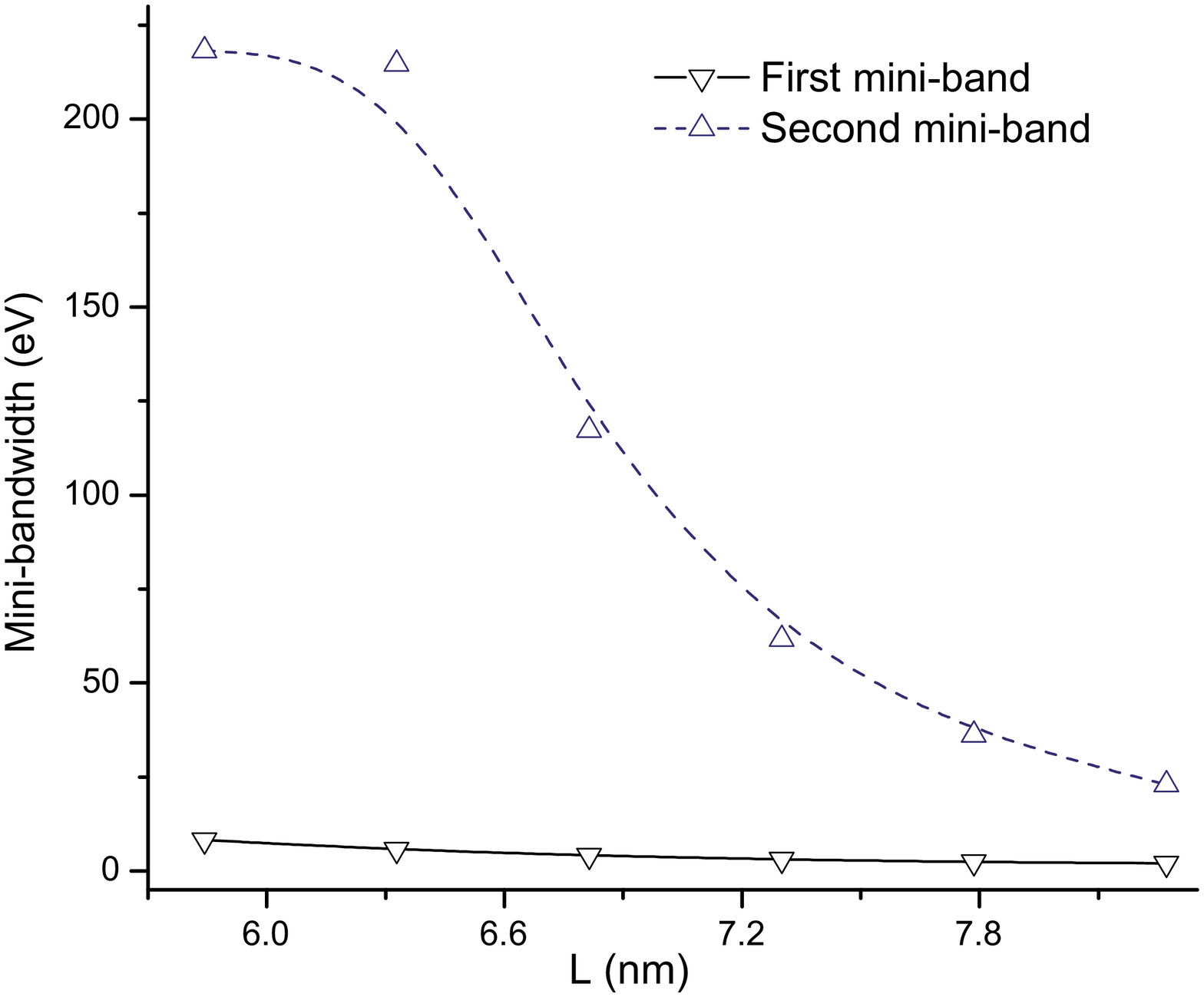} \,
\includegraphics[width=0.49\textwidth]{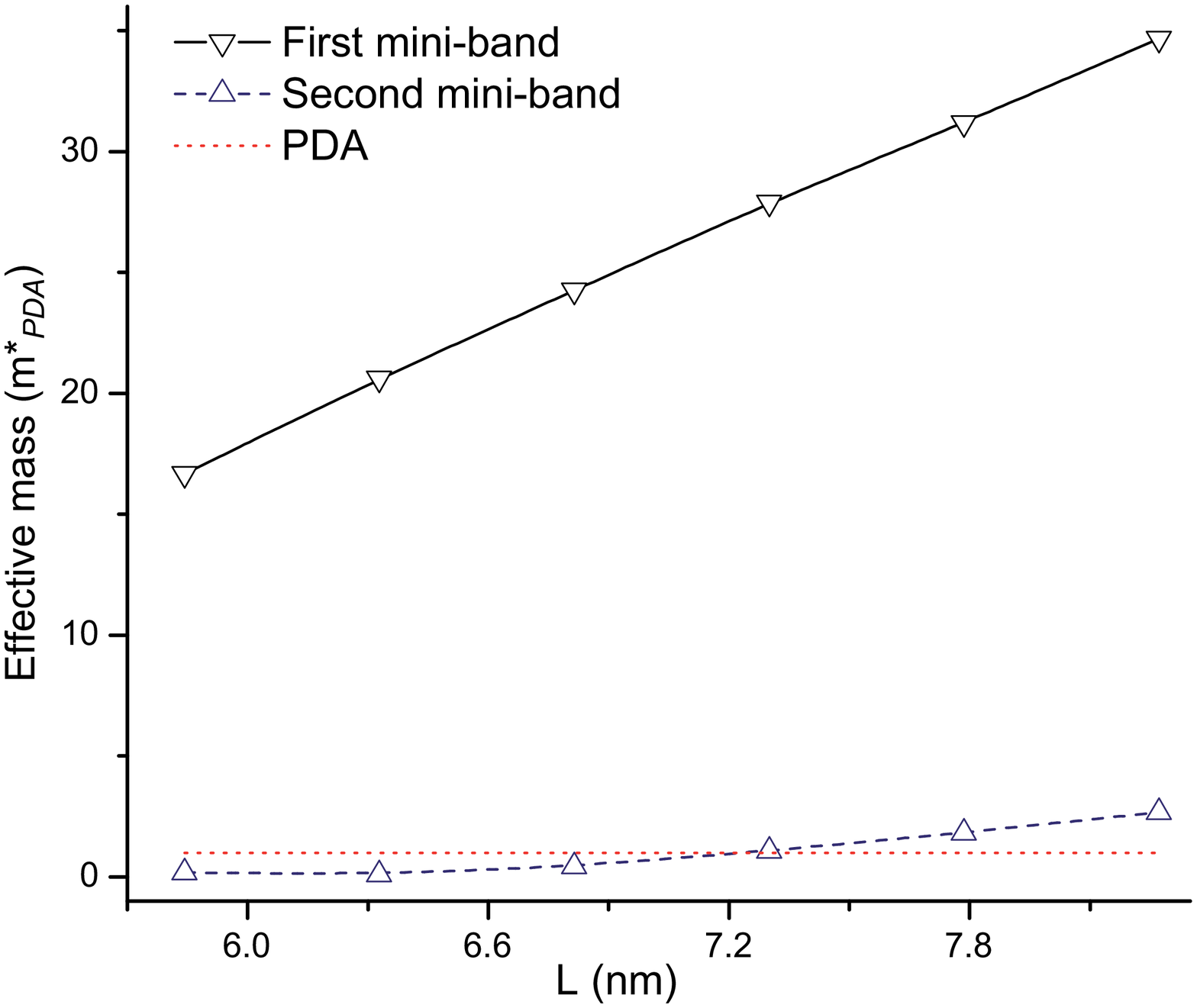}
\end{center}
\caption{\label{complete-8-y-8} \footnotesize{Infinite superlattice (PPP$_8$-PDA$_y$)$_{n}$. Left: mini-bandwidths. Right: mini-effective masses in units of the PDA electron effective mass (dotted line).}} 
\end{figure}

\noindent Let us finally examine the effect of increasing the number of monomers in the well (PDA) regions over the mini-bandstructure for the superlattices (PPP$_8$-PDA$_y$)$_{n \rightarrow \infty}$, with $y = 5,..,10$. The left part of Fig. \ref{complete-8-y-8} displays the results for the mini-bandwidths, where it is observed that these decrease as the width of the well grows. This can be explained by the fact that as the width of a single quantum dot increases the energies of its confined states decrease and the corresponding EMO's become more compact, thereby decreasing the overlap between adjacent levels in the superlattice.

\vspace{0.5cm}
\vspace{0.5cm}

\noindent The right part of Fig. \ref{complete-8-y-8} presents the results for the mini-effective masses. Again, it is seen that the effective mass of the second mini-band is largely insensitive to the barrier thickness. However, now it begins to grow at about $6.3$ nm and surpasses the PDA bulk value at about $7.3$ nm. This can be understood by realizing that as the well gets wider the states of the second mini-band steadily move further down from the edge of the barrier, steadily losing their PPP character and gaining more PDA character, which makes them sensitive to the well width. This also explains the appearance of a shoulder in the mini-bandwidth curve around $6.3$ nm, which must be due to the crossover from dominating PPP character to dominating PDA character.\\

\noindent The ECO approximation breaks down in the vicinity of abrupt heterojunctions \cite{inpreparation}. Therefore, the more similar the chemical structures of the parent polymers are and the longer the blocks of the superlattice become, so that the effect of the heterojunctions is effectively diminished, the better this approximation works \cite{inpreparation}. To assess the accuracy of the ECO method more quantitatively, a comparison against higher-level bandstructure calculations is necessary.

\section{Summary, Conclusions and Perspectives}
\label{sec:conclusions}

With the aid of the envelope molecular-orbital theory, molecular organic quantum-dot superlattices with the general structure 
(PDA$_x$-PPP$_y$)$_{n}$ were designed specifically to display two mini-bands with energies in between the conduction-band edges of the parent PPP and PDA polymers. Calibration curves of the gap between these two electron mini-bands and their effective masses as functions of the dot and barrier lengths were constructed, which allow tailoring of these heterostructures for specific electronic or optoelectronic applications, like infrared detectors and lasers. The mini-bandstructures were determined employing an envelope crystalline-orbital approximation with a Kronig-Penny model, which is seen to be a conceptually appealing and computationally very economical methodology.\\

\noindent As expected, it was found that the lower-lying mini-band exhibits a smaller bandwidth than the higher-lying one, and that these bandwidths decrease as the lengths of the dot and the barrier increase. Perhaps not so expectedly, it was also observed that the effective mass of the higher-lying mini-band is largely insensitive to increments in the dot and barrier lengths, remaining below the bulk PDA effective mass in most cases, whereas the effective mass of the lower-lying mini-band increases rapidly with such increments.\\

\noindent In future studies a quantitative comparison between the crystalline-orbital approximation and higher-level atomistic bandstructure methods will be performed. In addition, the full mini-bandstructure, i.e. including the hole minibands, will be considered, so that interband transitions and mini-excitonic effects in these molecular organic heterostructures can be determined. Furthermore, the effects of negative and positive doping of the superlattices, i.e. reduction and oxidation of the molecules, respectively, on their electronic, optical and conduction properties will be investigated.

\ack 
C.A.M.M. wishes to thank the organizing committee of QUITEL 2009 for partial financial support to participate in the conference. This research has been funded in part by Colciencias under contract 1106-45-221296.


\end{document}